\newcommand*{\ATLASLATEXPATH}{latex/}
\pdfoutput=1
\documentclass[cernpreprint, texlive=2016, UKenglish]{\ATLASLATEXPATH atlasdoc}
\usepackage [biblatex=true,backend=biber] {\ATLASLATEXPATH atlaspackage}
\usepackage{\ATLASLATEXPATH atlasbiblatex}

\usepackage{\ATLASLATEXPATH atlasphysics}
\graphicspath{{logos/}{figures/}}

\AtlasTitle{Search for resonant $WZ$ production in the fully leptonic final state in proton--proton collisions at $\mathbf{\sqrt{s} = 13}~\TeV$ with the ATLAS detector}

\AtlasAbstract{%
A search for a heavy resonance decaying into $WZ$ in the fully leptonic channel (electrons and muons) is performed.
 It is based on proton--proton collision data collected by the ATLAS experiment at the Large Hadron Collider at a centre-of-mass energy of 13~\TeV, corresponding to  an integrated luminosity of 36.1~fb$^{-1}$.
  No significant excess is observed over the Standard Model predictions and limits are set on the production cross section times branching ratio of a heavy vector particle produced either in quark--antiquark fusion or through vector-boson fusion. Constraints are also obtained on the mass and couplings of a singly charged Higgs boson, in the Georgi--Machacek model, produced through vector-boson fusion.
}

\author{The ATLAS Collaboration}

\PreprintIdNumber{CERN-EP-2018-077}

\AtlasJournalRef{Phys.\ Lett.\ B 787 (2018) 68}
\AtlasDOI{10.1016/j.physletb.2018.10.021}

\hypersetup{pdftitle={ATLAS document},pdfauthor={The ATLAS Collaboration}}

\begin{document}

\maketitle

%\tableofcontents

%-------------------------------------------------------------------------------
\section{Introduction}
\label{sec:intro}
\label{introduction}
Searches for diboson resonances provide an essential test of theories of electroweak symmetry breaking beyond the Standard Model (BSM). Vector resonances are predicted in various BSM scenarios, such as in extended gauge models~\cite{Langacker:1984dc,Altarelli:1989ff},
Little Higgs models~\cite{ArkaniHamed:2002qy}, 
Composite Higgs models and walking technicolor~\cite{Agashe:2004rs,Giudice:2007fh,Foadi:2007ue}, unitarized Electroweak Chiral Lagrangian models~\cite{Delgado:2017cls}, as well as in theories with extra dimensions~\cite{Randall:1999ee,Davoudiasl:1999tf,Csaki:2003dt}. In addition, new scalar diboson resonances result from models with an extended Higgs sector~\cite{deFlorian:2016spz,Ivanov:2017dad}.
This Letter reports on a search for a $WZ$ resonance in the fully leptonic decay channel $\ell\nu~\ell\ell$ ($\ell =e$ or $\mu$),  produced either by quark--antiquark ($q\bar{q}$) fusion or by vector-boson fusion (VBF). 
The proton--proton collision data were collected by the ATLAS detector~\cite{PERF-2007-01} at the Large Hadron Collider (LHC)  at a centre-of-mass energy $\sqrt{s}=13~\TeV$.

Parameterized Lagrangians~\cite{deBlas:2012qp,Pappadopulo:2014qza,Greco:2014aza} incorporating a heavy vector triplet (HVT) permit the interpretation of searches for vector resonances
in a generic way. Here, the simplified phenomenological Lagrangian of Ref.~\cite{Pappadopulo:2014qza} is used.  
The coupling of the new heavy vector resonance, $V$, to the Higgs boson and the Standard Model (SM) gauge bosons is parameterized by $g_Vc_H$ 
and to the fermions via the combination $(g^2/g_V)c_F$, where $g$ is the SM SU(2) gauge coupling. The parameter $g_V$ represents the typical strength of the vector-boson interaction, while the parameters $c_H$ and $c_F$ are expected to be of the order of unity in most models. The vector-boson scattering process, $pp \to V jj \to WZ jj$,  is only sensitive to the gauge boson coupling and, in this case, the benchmark model used to interpret the results assumes no coupling of the heavy vector resonance to fermions. 
	
The Georgi--Machacek model (GM)~\cite{Georgi:1985nv,Chanowitz:1985ug} is used as a benchmark for a singly charged scalar resonance. The model extends the Higgs sector by including one real and one complex triplet, while preserving custodial symmetry, ensuring that the parameter $\rho = M_W^2/(M_Z^2 \cos^2 \theta_W) = 1$ at tree level. It is less experimentally constrained~\cite{Godfrey:2010qb,Hartling:2014aga} than other models with higher isospin representations, such as Little Higgs models or Left--Right symmetric models~\cite{Mohapatra:1974hk}.
    A parameter $\sin \theta_H $, representing the mixing of the vacuum expectation values, determines the contribution of the triplets to the masses of the $W$ and $Z$ bosons.
 The ten physical scalar states are organized into different custodial multiplets: a  fiveplet $(H_5^{++},H_5^+,H_5^0,H_5^-,H_5^{--})$ which is fermiophobic but couples to $WZ$,  a  triplet,
 and two singlets, one of which is identified as the $125~\GeV$ SM Higgs boson. Assuming that the triplet states are heavier than the fiveplet scalars, $H_5$  can only be produced by vector-boson fusion and the cross section is proportional to $\sin^2 \theta_H$. The singly charged members of this fiveplet are the object of the present search in the VBF channel. For both models the intrinsic width of the resonance is below 4\%, which is lower than the experimental resolution in nearly all the parameter space explored in the present analysis.
 
 The VBF process ($pp \to WZ j j$) is characterized by the presence of two jets with a large rapidity gap resulting from quarks from which a vector boson has been radiated. The absence of this topology is interpreted as $q\bar q$ production, collectively referred to here as $q\bar q$. The spectrum of the reconstructed invariant mass of the $WZ$ resonance candidates is examined for localized excesses over the expected SM background. Results are provided for the VBF and $q\bar q$ categories separately, neglecting possible signal leakage between them. 

Early results from the Tevatron~\cite{Abazov:2010dj,Aaltonen:2011zq} have put limits on the mass of a $W'$ boson of an extended gauge model~\cite{Altarelli:1989ff} in the $WZ$ channel between 180 GeV and 690 GeV.
	The present analysis extends the search for resonant $WZ$ production beyond that in Run 1 $pp$ collision data at $\sqrt{{s}}=8~\TeV$ performed by the ATLAS~\cite{EXOT-2013-07} and CMS~\cite{Khachatryan:2014xja} collaborations.
     Each 
 collaboration has combined results~\cite{EXOT-2014-18,EXOT-2016-01,CERN-EP-2017-077} from searches for heavy $VV$ and $VH$ resonances ($V = W$ or $Z$) based on Run 1 data 
 and on partial Run 2 data at $\sqrt{s}=13~\TeV$ in the fully hadronic ($qqqq$), semileptonic ($\ell\nu qq, \ell\ell qq, \nu\nu qq$), and fully leptonic ($\ell\ell\ell\ell$, $\ell\nu\ell\ell$, $\ell\ell\nu\nu$) final states.   More recent results from $VV$ and $VH$ resonance searches with data at $\sqrt{s}=13~\TeV$ have been reported in Refs.~\cite{1710.07235,1708.09638,1708.04445,1707.06958,1607.05621,1707.01303,1708.05379,Sirunyan:2018hsl,Sirunyan:2018ivv,Sirunyan:2018iff}. The various decay channels generally differ in sensitivity in different mass regions. The fully leptonic channel, in spite of a lower branching ratio, is expected to be particularly sensitive to low-mass resonances as it has lower backgrounds.
     A recent search~\cite{Sirunyan:2017sbn} by the CMS Collaboration for a charged Higgs boson produced by vector-boson fusion and decaying into $WZ$ in the fully leptonic mode, using 15.2~fb$^{-1}$ of data collected at $\sqrt{s}=13~\TeV$, has yielded limits on the coupling parameter of the GM model, as a function of mass. Limits on the GM model have also been set, based on analyses of same-charge $WW$ production by CMS~\cite{Sirunyan:2017ret} and opposite-charge $WW$ production by ATLAS~\cite{1710.01123}, using data at $\sqrt{s}=13~\TeV$ with an integrated luminosity of 36.1~fb$^{-1}$.

%-------------------------------------------------------------------------------
%-------------------------------------------------------------------------------
\section{ATLAS detector}
\label{sec:detector}
\newcommand{\AtlasCoordFootnote}{%
ATLAS uses a right-handed coordinate system with its origin at the nominal interaction point (IP)
in the centre of the detector and the $z$-axis along the beam pipe.
The $x$-axis points from the IP to the centre of the LHC ring,
and the $y$-axis points upwards.
Cylindrical coordinates $(r,\phi)$ are used in the transverse plane, 
$\phi$ being the azimuthal angle around the $z$-axis.
The pseudorapidity is defined in terms of the polar angle $\theta$ as $\eta = -\ln\tan(\theta/2)$.
Angular distance is measured in units of $\Delta R \equiv \sqrt{(\Delta\eta)^{2} + (\Delta\phi)^{2}}$.}

The ATLAS detector at the LHC has a
 cylindrical geometry with a near $4\pi$ coverage in 
solid angle.\footnote{\AtlasCoordFootnote}
The inner detector (ID), consisting of silicon pixel, silicon microstrip and transition radiation detectors, is surrounded by a thin superconducting solenoid
providing a \SI{2}{\tesla} axial magnetic field. It allows precise reconstruction of tracks from charged particles and measurement of their momenta up to a pseudorapidity of $|\eta| = 2.5$.
High-granularity lead/liquid-argon (LAr) sampling electromagnetic and steel/scintillator-tile hadron calorimeters, at larger radius,  provide energy measurements in the central pseudorapidity range $|\eta| < 1.7$. In the endcap and forward regions, LAr calorimeters
for both the EM and hadronic energy measurements extend the region of angular acceptance up to $|\eta| = 4.9$.
 Outside the calorimeters, the muon spectrometer incorporates multiple layers of trigger and tracking chambers in a magnetic field produced by a system of superconducting toroid magnets, enabling an independent precise measurement of muon track momenta for $|\eta|<2.7$. 
The ATLAS trigger system consists of a hardware-based level-1 trigger followed by a software-based high-level trigger~\cite{1611.09661}.

%-------------------------------------------------------------------------------
%-------------------------------------------------------------------------------
\section{Data and Monte Carlo samples}
\label{sec:Data}
\label{Data}
The data used in this analysis were collected during 2015 and 2016 with the ATLAS detector in $pp$ collisions at a centre-of-mass energy of $13~\TeV$ at the LHC. The minimum bunch crossing interval is 25~ns, with a mean number of 23 additional interactions per bunch crossing. The events are required to have passed combinations of  single-electron or single-muon triggers. 
The transverse momentum threshold of the leptons in 2015 is 24 GeV for electrons and 20 GeV for muons satisfying a loose isolation requirement based only on ID track information. Due to the higher instantaneous luminosity in 2016 the trigger threshold was increased to 26 GeV for both electrons and muons and tighter isolation requirements were applied. Additional electron and muon triggers that do not include any isolation requirements with transverse momentum thresholds of \pT = 60 GeV and 50 GeV, and a single-electron trigger requiring \pT $>$ 120 GeV with less restrictive electron identification criteria are used to increase the selection efficiency which reaches close to 100\%.
Events are accepted only if quality criteria for detector and data conditions are satisfied. With these conditions, the available datasets correspond to an integrated luminosity of 36.1~fb$^{-1}$. 

Samples of simulated data were produced by Monte Carlo (MC) generators with the detector response obtained from 
the $\GEANT$4 toolkit~\cite{Agostinelli:2002hh,SOFT-2010-01}. For some samples, the calorimeter response is obtained from  a fast parameterized simulation~\cite{ATL-PHYS-PUB-2010-013}, instead of $\GEANT$4.  
Additional  simulated inelastic $pp$ collisions, generated with \textsc{Pythia 8.186}~\cite{Sjostrand:2007gs} with the A2 set of tuned parameters~\cite{ATL-PHYS-PUB-2012-003} and the \textsc{MSTW2008LO}~\cite{Martin:2009iq} parton distribution function (PDF), were overlaid in order to model both the in- and out-of-time effects from additional $pp$ collisions (pile-up) in the same and neighbouring bunch crossings. The mean number of pile-up events in the MC samples was set to reflect the conditions in the data.

For the HVT interpretation, $W'\rightarrow WZ$ samples were generated. Two benchmark models, provided in Ref.~\cite{Pappadopulo:2014qza}, are used. In Model A, weakly coupled vector resonances arise from an extension of the SM gauge group~\cite{Barger:1980ix} with an additional SU(2) symmetry group and the branching fractions to fermions and gauge bosons are comparable.
In Model B, 
the heavy vector triplet  is produced in a strongly coupled scenario, as in a Composite Higgs model~\cite{Contino:2011np} and fermionic couplings are suppressed.
The parameter $g_V$ was set to 1 for Model A and to 3 for Model B. For both models, the parameter $c_F$ is assumed to be the same for all types of fermions. Simulated signal samples for the HVT benchmark Model A were generated for masses of vector resonances ranging from $250~\GeV$ 
to $3~\TeV$ with \textsc{MadGraph\_aMC@NLO}~2.2.2~\cite{Alwall:2014hca}, using the model file provided by the authors in Ref.~\cite{HVT-web} with the NNPDF23LO~\cite{Ball:2012cx} PDF set.  They are hadronized with \textsc{Pythia 8.186}. For interpretation in terms of Model B, the Model A cross sections are simply scaled. This is justified since the width remains well below the experimental resolution and the angular distributions are the same for both models.

For the VBF production channel, HVT samples were generated with $g_{V} = 1$ for masses ranging from $250~\GeV$ 
to $2~\TeV$.
The coupling parameter $c_H$ was set to 1 and all other couplings of the heavy triplet, including $c_F$, were set to 0  in order to maximize the VBF contribution.
 A dijet invariant mass of at least $150~\GeV$ was required during event generation.

For the GM signal samples,  $pp \to H_5^\pm jj \rightarrow W^\pm Z j j$  were produced with  \textsc{MadGraph\_aMC@NLO} 2.2.2 for the mass range 200 to 900 GeV  in the $H_5$-plane defined in~\cite{Degrande:2015xnm}, compatible with present limits~\cite{Hartling:2014aga,Zaro:2015ika}, using \textsc{GMCALC}~\cite{Hartling:2014xma} and with $\sin \theta_H=0.5$. They were produced at leading order, but normalized to next-to-leading order according to Ref.~\cite{deFlorian:2016spz}, where
the cross sections and widths, which scale as $\sin^2 \theta_H$, are also given. For these samples, a minimum \pT of $15~\GeV$ ($10~\GeV$) for the jets (leptons) was required during event generation and the pseudorapidity must be in the range $|\eta| < 5$ for jets and $|\eta| < 2.7$ for leptons. 

The background sources in this analysis include processes with two or more electroweak gauge bosons, namely $VV$ and $VVV$ as well as processes with top quarks, such as $t\bar{t}$, $t\bar{t}V$, single top and
$tZ$, and processes with gauge bosons produced in association with jets or photons ($Z+j$ and $Z\gamma$ ). MC simulation is used to estimate the contribution from background processes with three or more prompt leptons
while data-driven techniques are used for the case of background processes with at least one misidentified or non-prompt lepton. Simulated events are used for cross checks and to assess the systematic uncertainties in these backgrounds.

The dominant $WZ$ SM background process of order ($\alpha^2  \alpha_s^2$) involving color-exchange diagrams, here referred to as QCD $WZ$, was modelled using \textsc{Sherpa}~2.2.2~\cite{Gleisberg:2008ta} at 
next-to-leading order (NLO), 
 and includes hard-scattering, parton shower, hadronization and the
underlying events. Up to three additional partons generated at tree level were merged with the parton shower.
In order to estimate an uncertainty due to the parton shower modelling, two alternative $WZ$ samples were produced using \textsc{Powheg-Box v2}~\cite{Alioli:2010xd} interfaced with \textsc{Pythia~8.186} and \textsc{Herwig++}~\cite{Bahr:2008pv,} respectively.
A sample of the purely electroweak process $WZjj \to \ell\nu~\ell\ell~jj$ (labelled $WZjj$) with a matrix-element $b$-quark veto (at zero order in $\alpha_\text{s}$)  was generated separately with \textsc{Sherpa} 2.2.2. 
Contributions from $WZjb \to \ell\nu~\ell\ell~bj$ (labelled $WZbj$) are included in the $tZ$ sample described below.
To estimate an uncertainty due to the parton shower modelling 
an alternative \textsc{Madgraph+Pythia~8} sample was produced. This \textsc{Madgraph} sample includes $b$-quarks in the initial state and was split to provide a sample without (with) a $b$-quark in the final state to model the $WZjj$ ($tZ+WZbj$) background.

Samples of $q\bar q\to ZZ \to 4\ell$ or  $q\bar q\to ZZ \to \ell\ell~\nu\nu$ were generated by \textsc{Powheg-Box v2} at NLO, interfaced to \textsc{Pythia~8.186}
and normalized to NNLO by $K$-factors evaluated in Ref.~\cite{Brucherseifer:2014ama}. The $gg\to ZZ$  and tribosons were generated with \textsc{Sherpa}~2.1.1. The $t\bar{t}V$ and $tZ$ processes were generated at LO using \textsc{Madgraph\_aMC@NLO}, interfaced with \textsc{Pythia~8.186} ($t\bar{t}V$) and \textsc{Pythia~6.428} ($tZ$). The $t\bar{t}V$ samples were normalized to NLO predictions~\cite{deFlorian:2016spz}.

Finally  samples of SM backgrounds with at least one misidentified or non-prompt lepton, including $Z\gamma$, $W\gamma$, Drell--Yan $Z\to \ell \ell,~W\to\ell\nu$ as well as top-pair and single-top were generated to assist in the fake/non-prompt lepton background estimate. Events with $Z\gamma$ and $W\gamma$ in the final state  were generated with \textsc{Sherpa}~2.1.1.
Drell--Yan $Z\to \ell \ell,~W\to\ell\nu$ as well as top-pair and single-top production channels were generated with \textsc{Powheg-Box v2} and hadronized with \textsc{Pythia}. To avoid double counting the $Z\gamma$ events,  $Z$ events produced by the Drell--Yan process with a photon from final-state radiation with $\pt>10~\GeV$  were removed. The parton shower for processes with top quarks was modelled with \textsc{Pythia~6.428}. \textsc{Madgraph\_aMC@NLO} and \textsc{Pythia~8.186} were used for background processes involving a pair of top quarks accompanied by a $W$ boson or by a pair of charged leptons. The $Z$ and single-top cross sections were normalized to NNLO by $K$-factors evaluated in Ref.~\cite{Anastasiou:2003ds,Brucherseifer:2014ama}. 

SM backgrounds with Higgs bosons ($H, t\bar tH, VH$) contribute less than 0.1\% of the total background because of the low cross section and the requirement of a well reconstructed $Z$ boson decaying leptonically. These backgrounds are neglected.
 
%-------------------------------------------------------------------------------
%-------------------------------------------------------------------------------
\section{Reconstructed objects}
\label{sec:Objects}
\label{Objects}

Events are required to have at least one primary vertex with at least two associated tracks, each with transverse momentum 
$\pT > 0.4~\GeV$. If there is more than one vertex reconstructed in the event, the one with the largest track $\sum\pT^2$ is chosen as the hard-scatter primary vertex and is subsequently  used for the reconstruction of electrons, muons, jets and missing transverse momentum. 

Electron candidates are reconstructed from energy deposits in the EM calorimeter which are matched
to a well-reconstructed ID track originating from the primary vertex. 
The electron  identification is based on a likelihood evaluated from a multivariate discriminant. 
They are categorized as satisfying the \textit{medium} or the \textit{tight} reconstruction quality requirements, as defined in Ref.~\cite{ATLAS-CONF-2016-024}.  Only electrons with transverse  energy $\et > 25~\GeV$ in the pseudorapidity range $|\eta|<2.47$ are considered in this analysis.
The candidate electrons are required to pass an isolation condition: an upper value of the scalar sum of the transverse momentum of the tracks with $p_{\textrm T}  > 0.4~\GeV$ in a cone of size $\Delta R =\min(0.2,10~\GeV/\eT)$ around the electron, excluding the track of the electron itself, is chosen such that the efficiency is constant at 99\% for electrons in $Z \to ee$ events.
 For \textit{tight} electrons, an isolation requirement is imposed, based on calorimeter as well as track variables, which varies 
 as a function of transverse energy and yields an efficiency between 95\% and 99\% for electrons with \pT in the range 
 25--60~\GeV.
 For a pair of electrons sharing the same ID-track, the electron with higher cluster $\et$ is kept. 

 Muons are reconstructed by combining tracks from the inner detector  with tracks from the muon spectrometer. They are required to satisfy \textit{medium} or \textit{tight} quality requirements, as defined in Ref.~\cite{PERF-2015-10}. 
  Only muons with $p_{\textrm T}  > 25~\GeV$ and $|\eta| < 2.7$ are considered in this analysis. 
  Isolation requirements are also applied to all muons, based on the ratio $p_{\textrm T} ^{\textrm {varcone}}/p_{\textrm T}^\mu $, where $p_{\textrm T}^{\textrm {varcone}}$ is the scalar sum of the transverse momenta of the tracks with $p_{\textrm T}  > 1~\GeV$ in a cone of size $\Delta R = \min (10~\GeV/p_{\textrm T}^\mu,0.3)$ around   the muon, excluding the muon track itself. This isolation gives 99\% efficiency, independently of $\eta$ or $p^\mu_{\textrm T} $, in $Z\to\mu\mu$ samples.  

Electron and muon candidates are required to originate from the primary vertex. Thus, the significance of the track's transverse impact parameter calculated relative to the beam line, $|d_0/\sigma_{d_0}|$, must be less than three for muons and less than five for electrons, and the longitudinal impact parameter, $z_0$ (the difference between the value of $z$ of the point on the track at which $d_0$ is defined and the longitudinal position of the primary vertex), is required to satisfy $|z_0 \cdot \sin(\theta)| < 0.5$~mm.

Jets are reconstructed from clusters of energy deposition in the calorimeter~\cite{ATLAS-CONF-2013-004} using the anti-$k_t$ algorithm~\cite{Cacciari:2008gp} with a radius parameter $R = 0.4$. Events with jets arising from detector noise or other non-collision sources are discarded~\cite{PERF-2011-03}. 
This search considers jets with $\pT > 30~\GeV$ in the range $|\eta|<4.5$.
 Furthermore, to mitigate the pile-up contamination, a jet vertex tagger~\cite{Aad:2015ina}, based on information about tracks associated with the primary vertex and pile-up vertices, is applied to jets with $\pT < 60~\GeV$ and $|\eta| <$~2.4. The selected working point provides at least 92\% efficiency.
The energy of each jet is calibrated and corrected for detector effects using a combination of simulated events and in situ methods in $13~\TeV$ data~\cite{ATL-PHYS-PUB-2015-015}.

  As lepton and jet candidates can be reconstructed from the same detector information, a procedure to resolve overlap ambiguities is applied. If an electron and a muon share the same ID track, the muon is selected.  Reconstructed jets which overlap with electrons or muons in a cone of size $\Delta R = 0.2$ are removed.

Jets containing $b$-hadrons are identified as \textit {b-jets} by the MV2c10 $b$-tagging algorithm~\cite{PERF-2012-04}, which uses information such as track impact-parameter significances and positions of explicitly reconstructed secondary decay vertices. A working point corresponding to 85\% $b$-tagging efficiency on a sample of $t\bar t$ events is chosen~\cite{Aaboud:2018xwy}, with a light-flavour jet rejection factor of about 34 and a $c$-jet rejection of about 3. Correction factors are applied to the simulated event samples to compensate for differences between data and simulation in the $b$-tagging efficiency for $b$-jets, $c$-jets and light-flavour jets. 
    
The missing transverse momentum, $\vec{p}_{\textrm T}^{\textrm {miss}}$, and its magnitude \MET, are calculated from the imbalance in the sum of visible transverse momenta of reconstructed physics objects: electrons, muons and jets, as well as a ``soft'' term reconstructed from tracks compatible with the primary vertex and not associated with any of those objects~\cite{ATL-PHYS-PUB-2015-027,ATL-PHYS-PUB-2015-023}.

%-------------------------------------------------------------------------------
\section{Event selection}
\label{sec:Selection}
 \label{Selection}
Preselection criteria are first applied to all the event samples. The presence of three prompt leptons is required, two of which will be associated with the $Z$ boson, and are required to satisfy the \textit{medium} quality requirement (Section~\ref{Objects}). The $Z$ boson candidate is reconstructed from the two leptons of same flavour and opposite charge, whose invariant mass is closest to the on-shell mass $m_Z$, and in the range $|m_{\ell\ell} - m_Z| < 20~\GeV$. The third lepton, associated with the $W$ boson decay, is required to satisfy the \textit {tight} quality criteria to enhance the background rejection. To ensure that the trigger efficiency is well determined, at least one of the candidate leptons is required to have $\pt>27~\GeV$.

To suppress background processes with at least four prompt leptons, events with a fourth lepton candidate satisfying looser selection criteria are rejected. For this looser selection, the requirement on the minimum \pT of the leptons is lowered to $\pT>7~\GeV$ and \textit {medium} identification requirements are used for both the electrons and muons. 

 Since there is a neutrino in the signal events, $\met>25~\GeV$ is also required.
 The third lepton and the missing transverse momentum are assumed to result from the $W$ boson decay. The longitudinal momentum $p_z^\nu$ of the neutrino is calculated by requiring  that the invariant mass  of the lepton--neutrino system be equal to the $W$ mass. 
The solution results in a quadratic equation which leads to two possible solutions. If they are real, the one with the smaller $|p_z^\nu|$ is chosen since it was found to provide a better agreement with the truth. Otherwise, the real part is chosen. The invariant mass, $m_{WZ}$, of the $WZ$ resonance candidate is then reconstructed using the chosen solution for $p_z^\nu$ along with the four-momenta of the three charged leptons.

\begin{figure}[!htbp]
\begin{center} 
\subfloat[ ]{\includegraphics[width=.50\textwidth]{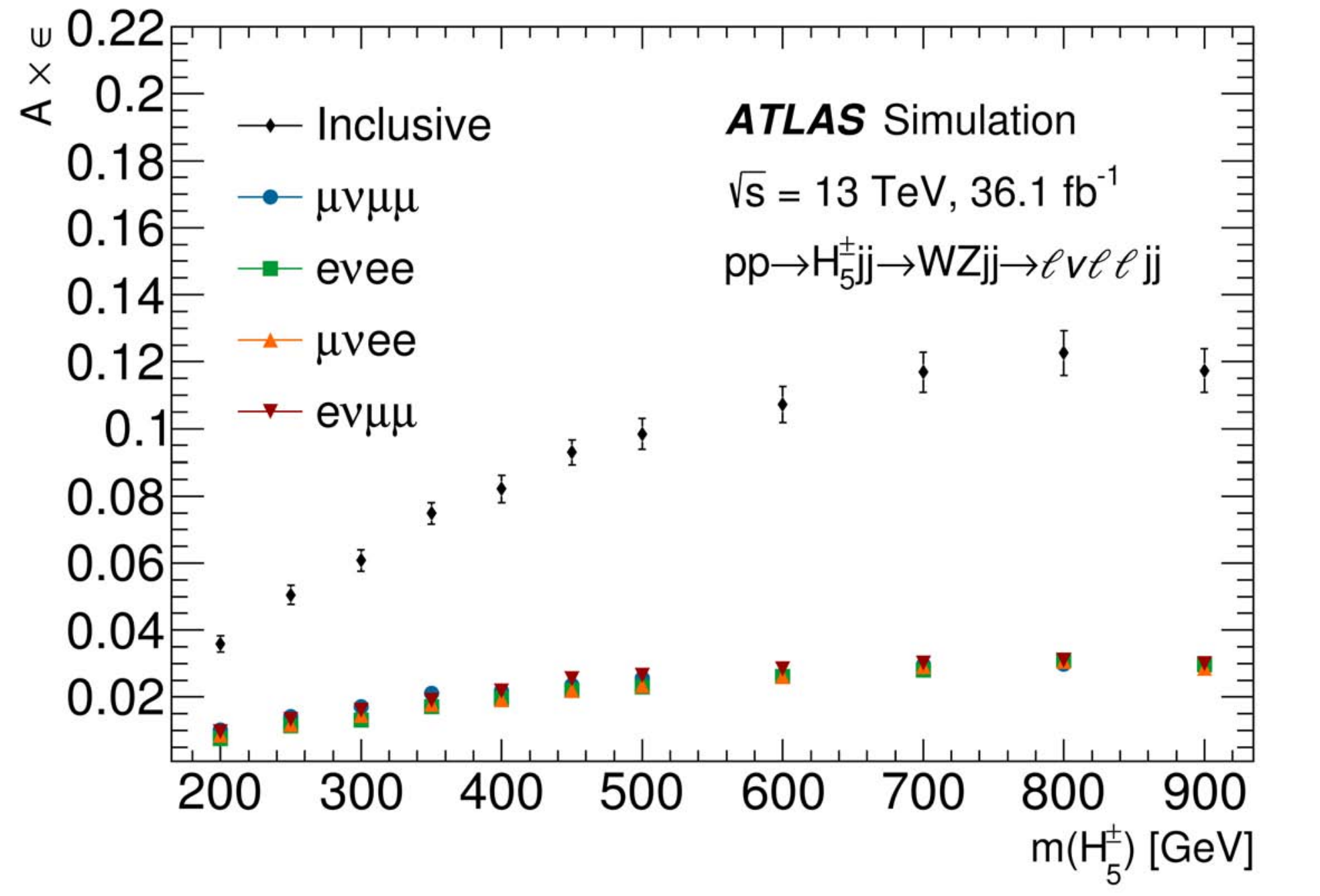}\label{fig:AxE:Hp5AC}}
\subfloat[ ]{\includegraphics[width=.50\textwidth]{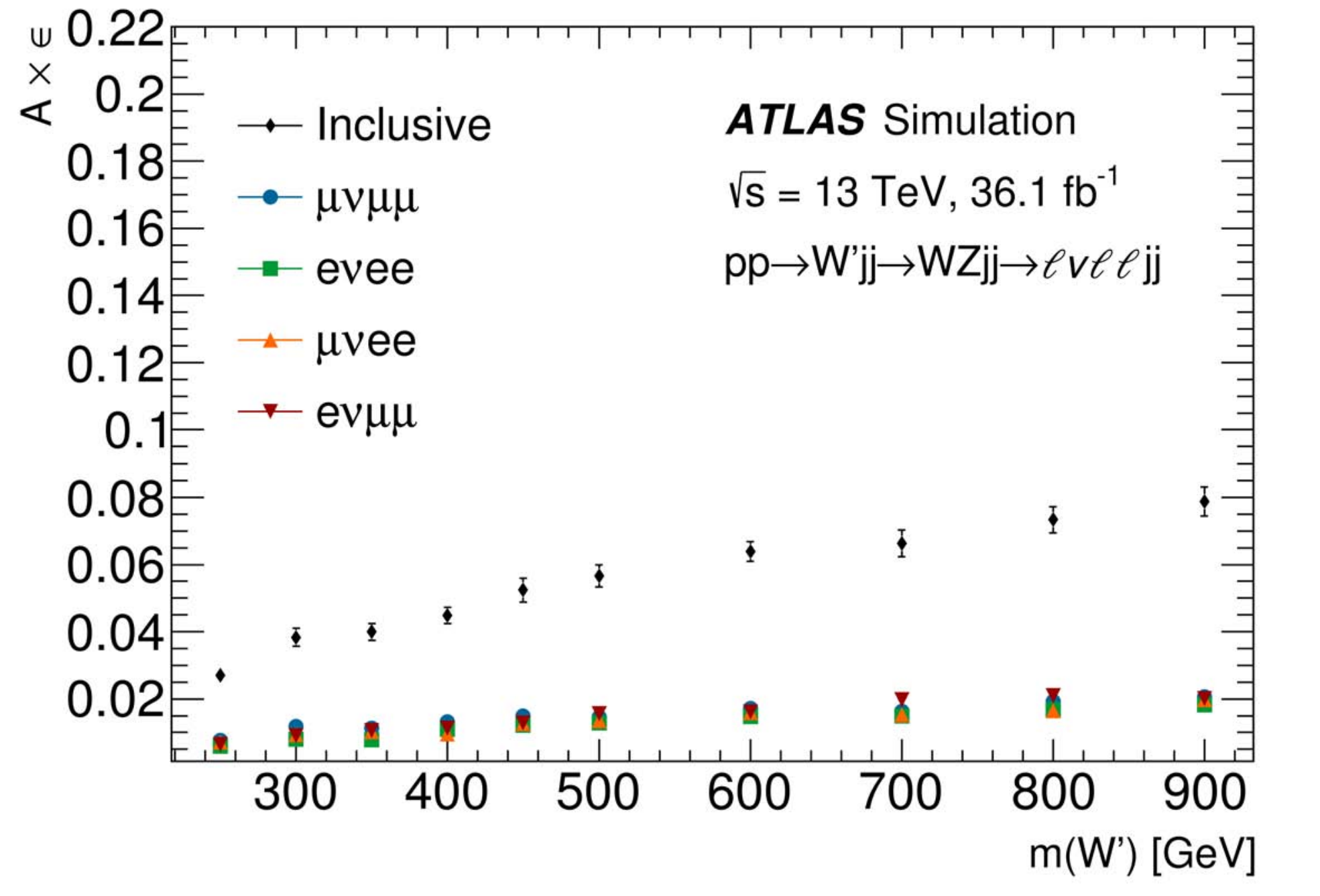}\label{fig:AxE:HVTACVBS}}

\end{center}
 \caption{The signal selection acceptance times efficiency  (A$\times \epsilon$), defined as the ratio of the number of MC signal events in the VBF category to the number of generated signal events, is presented as a function of the resonance mass. Figure~\ref{fig:AxE:Hp5AC} corresponds to the GM Model ${H_{5}^{\pm}}$ while Figure~\ref{fig:AxE:HVTACVBS} corresponds to the HVT models. The  A$\times \epsilon$ is shown for each decay channel and the sum of all lepton flavour combinations (inclusive). The error bars shown in each figure represent the total statistical and systematic uncertainties.}
\label{fig:AxE:VBF}
\end{figure}

The selected events are then separated into two categories targeting different production mechanisms: VBF and $q\bar{q}$. 
The VBF category contains events with two or more jets with $\pt>30~\GeV$   which fail the $b$-tagging requirements described in Section~\ref{sec:Objects}.  The dijet pair defined by the two highest-\pt jets in the event must also have large $\eta$ separation ($|\Delta \eta_{jj}| >3.5$) and  an invariant $m_{jj}$ above $500~\GeV$.
If more than two jets are found in an event, the two highest-$p_{\textrm T}$ jets are considered. 
By imposing a $b$-jet veto, backgrounds containing one or more top quarks, including $t\bar t$, $t\bar{t}+W/Z$,  and $tZ$ are suppressed. 

The net acceptance times efficiency (A$\times\epsilon$) of the selection, relative to signal events generated for $H_5^{\pm}$ and HVT models in the VBF category is shown in Figure~\ref{fig:AxE:VBF}. 
The generation of the GM Model  ${H_{5}^{\pm}}$ events had the following requirements:  \pT(jets) > 15~\GeV, \pT(leptons) > 10~\GeV, $|\eta|$(jets) $< 5$ and $|\eta|$(leptons)~$< 2.7$. Decays of $W$ bosons into all lepton flavours, and of $Z$ bosons into $e^+e^-$ and $\mu^+\mu^-$, were simulated. The $Z \to \tau^+\tau^-$ decays give a negligible contribution and were not included in the simulation, but the A$\times \epsilon$ shown was scaled to include all decays.  For the HVT VBF samples, $m_{jj} > 150~\GeV$ was required at generator level. Decays of $W$ and $Z$ bosons into all flavours of leptons were included. 
For HVT and $H_5^{\pm}$ the A$\times\epsilon$ falls in the range 2--8\%  and 3--12\% respectively for resonance masses ranging between 200 and 900~\GeV, the difference being due, with approximately equal importance, to the generator level selection and to the different angular distributions of the final products.

\begin{figure}[!htbp]
\begin{center}
\includegraphics[width=.50\textwidth]{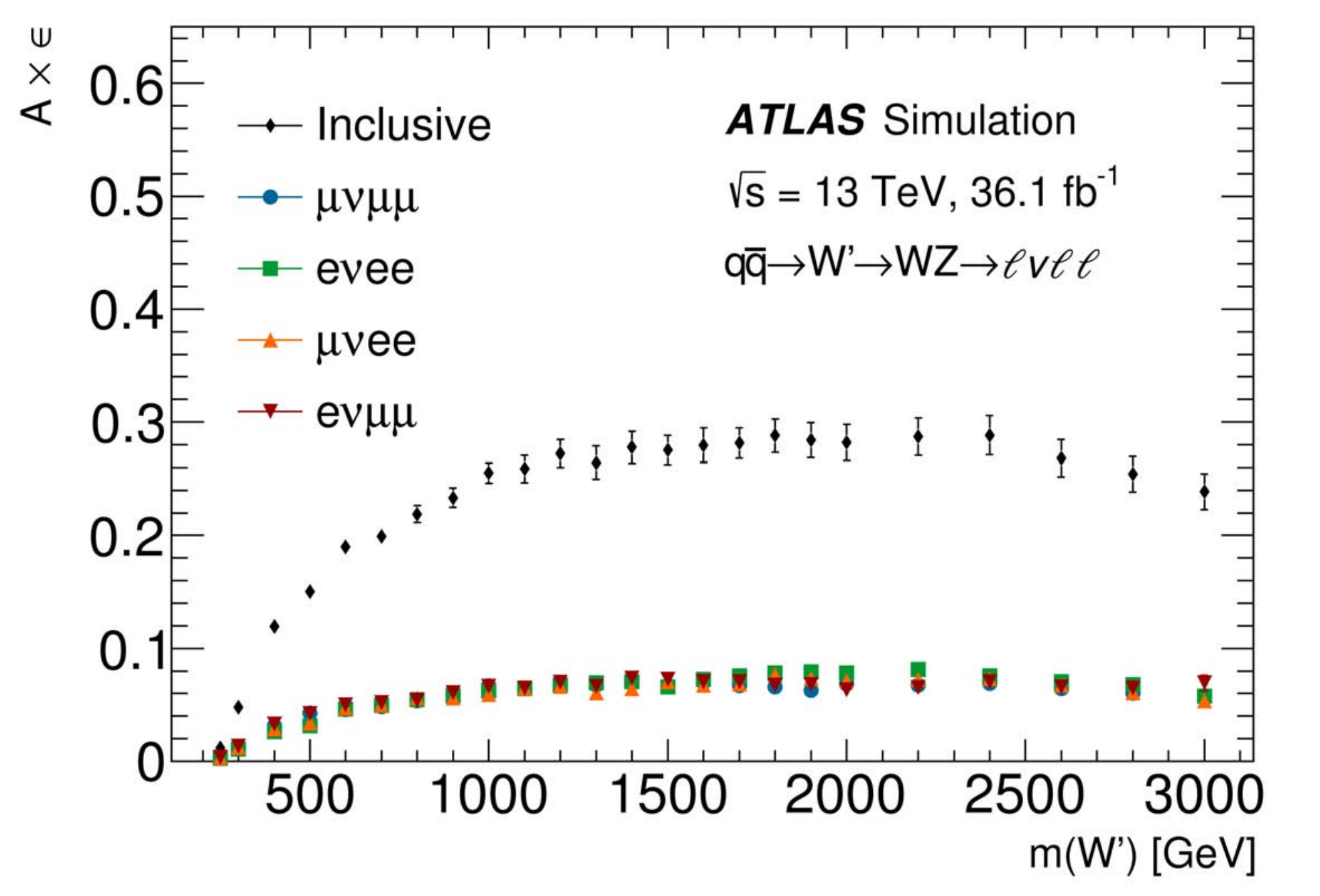}
\end{center}
 \caption{The signal selection acceptance times efficiency  (A$\times \epsilon$), defined as the ratio of the number of MC signal events in the $q\bar q$ category to the number of generated signal events, as a function of the HVT resonance mass.   The error bars represent the total statistical and systematic uncertainties. }
\label{fig:AxE:HVTACqq}
\end{figure}

The remaining  events are assigned to the $q\bar q$ category signal region. For this category, the $W$ and $Z$ bosons from a resonance produced in the s-channel with $m_{WZ}$ larger than $250~\GeV$  are expected to have transverse momenta close to 50\% of its mass. The requirements $p_{\textrm T}^{W}/m_{WZ} > 0.35$ and $p_{\textrm T}^{Z}/m_{WZ} > 0.35$ enhance the sensitivity to the signal. 
The overall selection efficiency relative to generated event increases from about 15\%  to 25\% for resonances masses ranging from $500~\GeV$ to $3~\TeV$ as illustrated in Figure~\ref{fig:AxE:HVTACqq}. Decays of $W$ and $Z$ bosons into all flavours of leptons are included at event generation.
The A$\times \epsilon$ values decrease for resonance masses above approximately $2~\TeV$ due to the collinearity of electrons from the $Z\rightarrow ee$ decays which spoils the isolation.

%-------------------------------------------------------------------------------
%-------------------------------------------------------------------------------
\section{Background estimation}
\label{sec:Background}
\label{Background}

\begin{figure}[!htbp]
\begin{center}
\subfloat[ ]{\includegraphics[width=.5\textwidth]{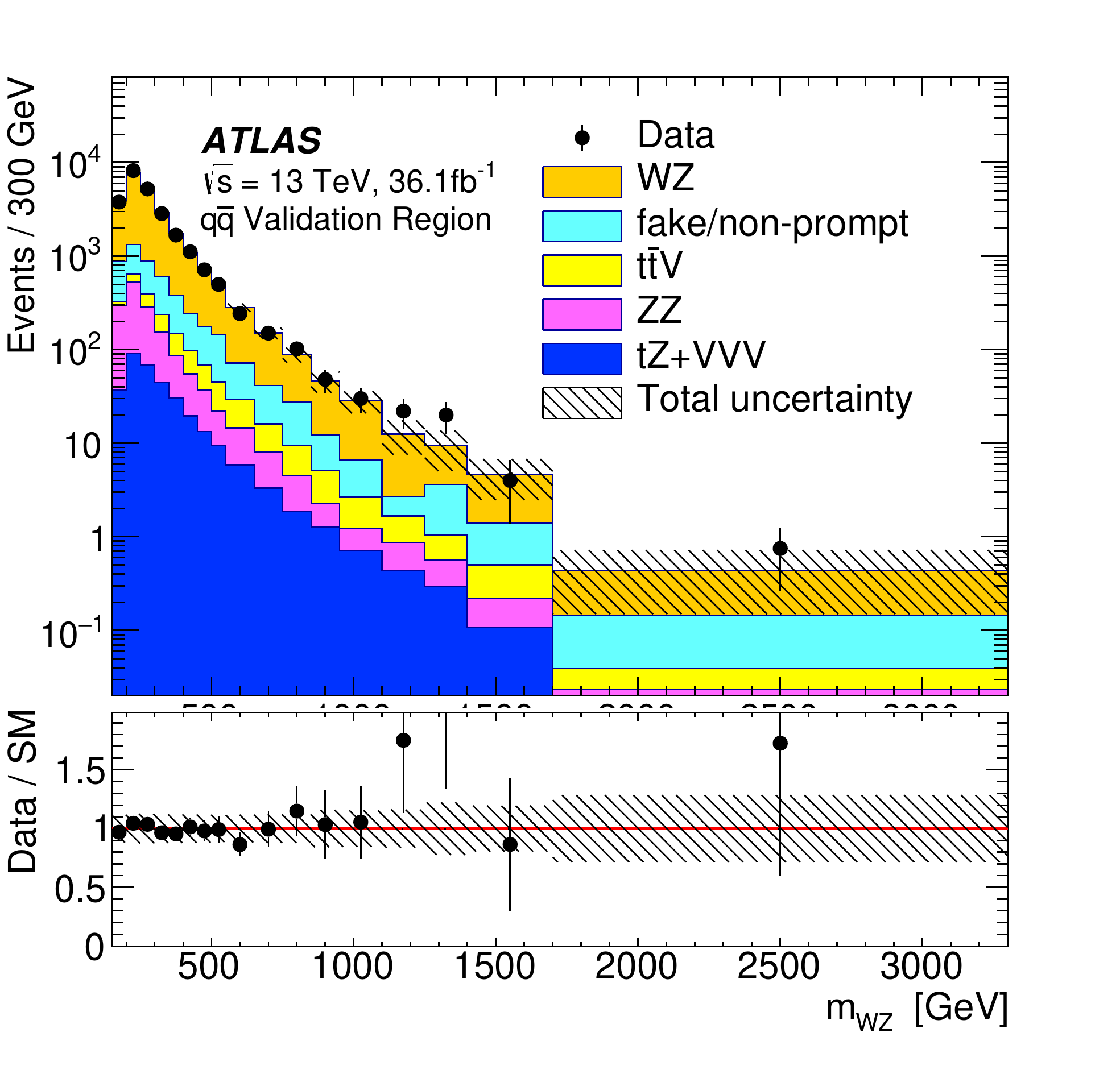}\label{fig:Background:WZDistVRa}}
\subfloat[ ]{\includegraphics[width=.5\textwidth]{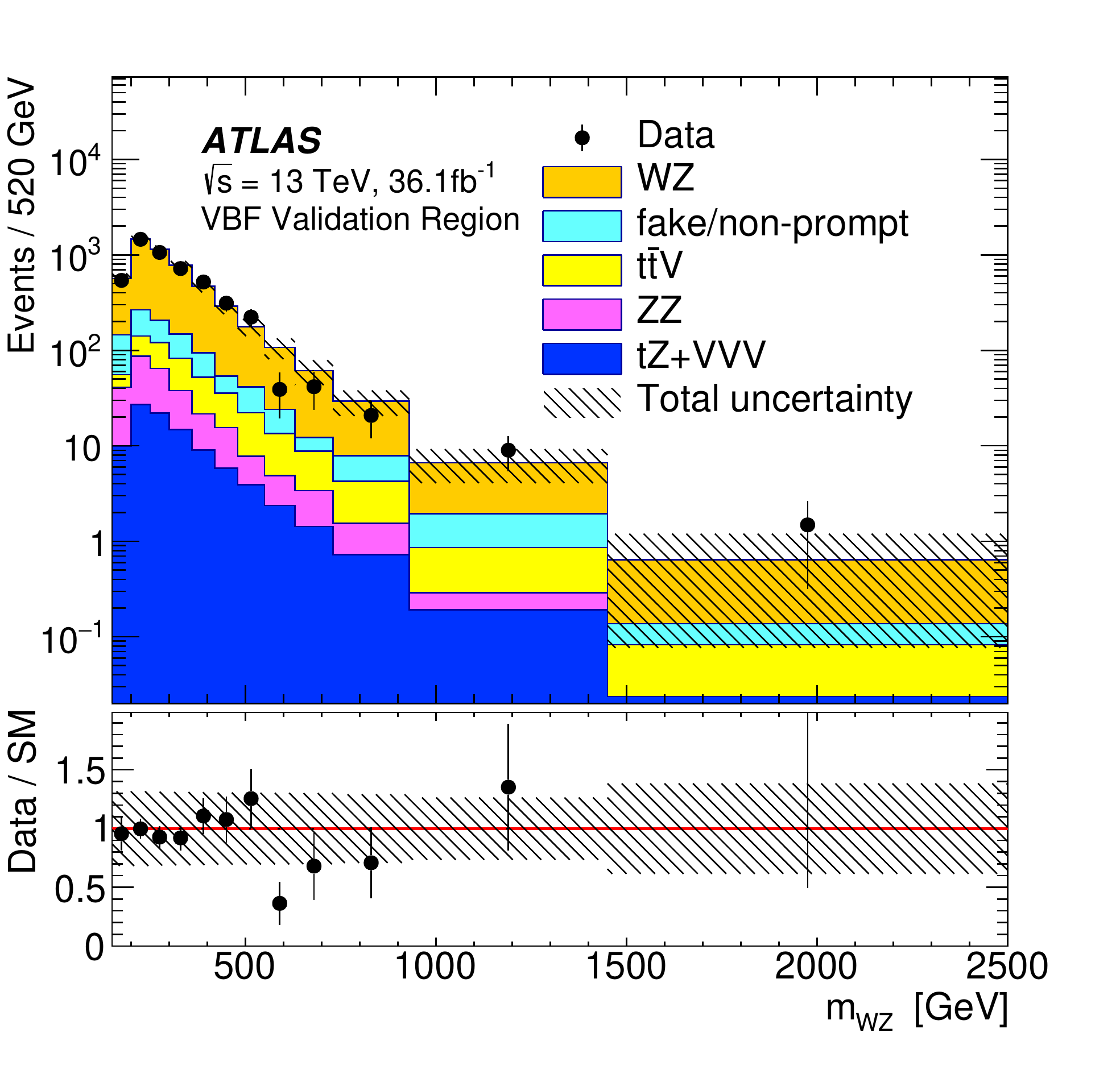}\label{fig:Background:WZDistVRb}}
\end{center}
 \caption{Observed and expected distributions of the $WZ$ invariant mass in \protect\subref{fig:Background:WZDistVRa} the $q\bar{q}$ validation region and \protect\subref{fig:Background:WZDistVRb} the VBF validation region. The points correspond to the data and the histograms to the expectations for the different SM processes. 
 The uncertainty in the total background prediction, shown as bands, combines statistical, theory and systematic contributions.The bin sizes were chosen to reflect the resolution of a reconstructed signal, and  the width of the previous to last bin is used to scale the bin contents. The last bin contains the overflow.
\\
 }
\label{fig:Background:WZDistVR}
\end{figure}

The dominant background in the 
resonance search is the SM production of $WZ$. Its normalization and shape are estimated from MC and validated in dedicated validation regions by comparing the data and MC distributions.
 Events in the validation regions are selected in exactly the same way as those in their corresponding signal categories except for the following requirements. 
 The VBF $WZ$ validation region is defined by inverting the requirements on the dijet variables: $100<m_{jj}<500~\GeV$ and  $|\Delta\eta_{jj}|<3.5$. The $WZ$ $q\bar{q}$ validation region requires the events to have $p_{\textrm T}^{\textrm Z}/m_{WZ}<0.35$ or $p_{\textrm T}^{W}/m_{WZ}<0.35 $.  
These validation regions are dominated by the $WZ$ contribution, with a purity higher than 80\%. For the benchmark models with parameters given in Section~\ref{Data}, the signal contamination in the $q\bar q$ (VBF) validation region is below 5\% (1\%).
The reconstructed $m_{WZ}$ mass in the validation regions is shown in Figure~\ref{fig:Background:WZDistVR}, where good agreement of data with the background prediction is observed.

Events from $Z+$jets, $Z\gamma$, $W\gamma$, \ttbar, single top or $WW$ where jets or photons were misidentified as leptons (here called \textit {fake/non-prompt} leptons),  can also satisfy the selection criteria. The distribution shapes and number of fake/non-prompt lepton events are estimated for both the $q\bar{q}$ and VBF categories by a data-driven method using a global matrix which exploits differences in object characteristics between real and fake/non-prompt leptons on a statistical basis. Details of the method, here referred to as ``Matrix Method'', can be found in Ref.~\cite{STDM-2014-02}.

Other backgrounds include $t\bar t V$, $ZZ$, $tZ$, $WZbj$ and triple boson production. They are estimated by Monte Carlo simulation (Section~\ref{Data}). The $tZ$, $WZbj$ and $VVV$ backgrounds are added as a single contribution, here called $tZ$+$VVV$.

%-------------------------------------------------------------------------------
%-------------------------------------------------------------------------------
\section{Systematic uncertainties}
\label{sec:Systematic}
Systematic uncertainties result  from the theoretical modelling of backgrounds and from object and event reconstruction. 

The uncertainty in the normalization 
of the \textsc{Sherpa} samples of SM  $WZ$ background  is evaluated by taking into account
the variations obtained with different PDF sets~\cite{Buckley:2014ana}. The nominal  set NNPDF30nloas0118 is compared with other samples generated with the CT14nnlo and MMHT2014nlo68cl PDF sets and the uncertainty is evaluated from the maximum differences. It is estimated to be below 6\% in all mass bins for both the VBF and $q\bar{q}$ categories. The uncertainty associated with the choice of renormalization  and factorization scales, $\mu_\text{R}$ and $\mu_\text{F}$,  is taken as the maximum downward and upward variation when the scales are varied independently  by factors of 1/2 and 2. 
While these uncertainties can in principle affect the shape of the $m_{WZ}$ distribution, in practice the shape differences do not have a strong impact on the sensitivity of the search. The uncertainties are therefore treated as normalization uncertainties, taken to be
20\% and 40\%  respectively. Shape systematic uncertainties associated with showering and hadronization of the QCD $WZ$ are evaluated by comparing the \textsc{Powheg-Box v2} samples interfaced with \textsc{Pythia~8.186} and \textsc{Herwig}. For the electroweak $WZ$ process  the \textsc{Sherpa}~2.2.2 and \textsc{Madgraph+Pythia~8} predictions are compared. 
This uncertainty band ranges from 10\% to 30\% for both categories.

The uncertainties assigned to the cross sections of the other  background sources consist of a contribution from PDF uncertainties and from QCD scale uncertainties. 
They are estimated to be 10\% for $ZZ$, 13\% for $ttV$, 20\% for $VVV$ and 15\% for $tZ$. 

The theoretical uncertainties in the cross section and acceptance of the simulated signal samples are evaluated in a similar way to the background. The PDF errors are taken from the NNPDF LO PDF error set, and the NNPDF set is also compared with the CTEQ6L1 and MSTW2008lo68cl sets. The different predictions from these PDF sets are taken as an extra contribution to the overall uncertainty.    
For both the $q\bar q$ and VBF categories, the uncertainties are typically below 5\%.
This procedure was followed for each mass point and a generator-level event selection was chosen to closely mimic the one used in the reconstruction-level analysis. Scale and PDF  uncertainties are not correlated between signal and background.

An  uncertainty due to the reconstruction efficiency, momentum scales and resolution of electrons and muons is evaluated by varying correction factors applied to the MC samples~\cite{Aaboud:2016vfy,PERF-2015-10} within appropriate limits. 

The jet energy scale and resolution uncertainties~\cite{PERF-2011-03} are also taken into account as they affect the shape and normalization of the background distributions. The uncertainty due to  $b$-tagging~\cite{ATL-PHYS-PUB-2016-012} is also included. 

Missing transverse momentum is calculated using the preselected leptons, jets and other reconstructed objects. The uncertainties in the reconstruction of those objects are then used to evaluate the uncertainty in  \met reconstruction.
Those due to the $\pt$ scale and resolution of the soft term are also considered~\cite{ATL-PHYS-PUB-2015-027,ATL-PHYS-PUB-2015-023}. 

An uncertainty in the prediction of the fake/non-prompt background is also taken into account as it affects the shape and normalization of the background distributions.
The total uncertainty is about 20\% (27\%) for the $q\bar{q}$ (VBF) category. It is slightly larger for the VBF category because of the higher statistical uncertainty derived from the Matrix Method (Section~\ref{Background}).

The uncertainty in the integrated luminosity is 2.1\%. It is derived, following a methodology similar to the one detailed in Ref.~\cite{DAPR-2013-01}, from a calibration of the luminosity scale using $x$--$y$ beam-separation scans performed in August 2015 and May 2016.

%-------------------------------------------------------------------------------
%-------------------------------------------------------------------------------
\section{Results}
\label{sec:result}
The $WZ$ invariant mass distribution, $m_{WZ}$, obtained as the sum of all four lepton-flavor permutations,  is used as the discriminating variable, with bin widths comparable to the expected resolution of a narrow resonant signal. 
A binned likelihood function, constructed from the Poisson probability of the sum, in each bin, of the contributions of the background and of a hypothetical signal of strength $\mu$ relative to the benchmark model, is used to set limits on the presence of a signal.
The fit is performed in the signal region for the $q\bar{q}$ and VBF categories separately. 
The systematic uncertainties described above (Section~\ref{sec:Systematic}) enter as nuisance parameters with Gaussian or log-normal prior distributions, in convolution with the nominal background distribution. 

The effects of systematic uncertainties are studied for hypothesized signals using the signal-strength parameter $\mu$.
The list of leading sources of uncertainty in the 95\% confidence level (CL) upper limit on the $\mu$ value is given in Table~\ref{tab:NPranks} together with their relative importance ($\Delta \mu / \mu$). The values are quoted separately
for a hypothetical HVT signal of mass $m(W') = 800~\GeV$ in the $q \bar q$ category and a GM signal of mass $m(H_{5}^{\pm})=450~\GeV$ in the VBF category.
Apart from the statistical uncertainties in the data, the uncertainty with the largest impact on the sensitivity of the searches is related to the $WZ$ background modelling. 

\begin{table}[]
\centering
\vspace{7pt}
%\label{tab:NPranks}
\begin{tabular}{l|c|c}
\hline
\hline
Source & \multicolumn{2}{c}{$\Delta \mu / \mu$ [\%] }\\
\hline
 &$q\bar q$ Category &        VBF Category  \\

 &$m(W')=800\ \gev$ &       $m(H_{5}^{\pm})=450\ \gev $  \\

\hline
$WZ$ background modelling : Scale, PDF                    & 5     &  11    \\
$WZ$ background modelling : Parton Shower                 & 10    &   6 \\ 
MC statistical uncertainty  	               &  7    &  8   \\ 
Electron identification 		       &  4    &  2  \\ 
Muon identification                            &  3    &  3   \\  
Jet uncertainty                                &  1    &   8   \\ 
Missing transverse momentum                &  2    &   1 \\     
Fake/non-prompt                                          &  1    &   5 \\                                            
 \hline
Total systematic uncertainty                   & 17   & 21  \\
Statistical uncertainty                        & 53   & 52  \\
\hline \hline
\end{tabular}

\caption{
Impact of the dominant sources of relative uncertainties on the 95\% CL upper limits of the signal-strength parameter ($\mu$) for a hypothetical HVT signal of mass $m(W') = 800~\GeV$ in the $q \bar q$ category and a GM signal of mass $m(H_{5}^{\pm}) = 450~\GeV$ in the VBF category. The effect of the statistical uncertainty on the signal and background  samples is also shown. Sources of systematic uncertainty with an impact of less than 2\% in both categories are not shown.}
\label{tab:NPranks}
\end{table}

The numbers of background events are extracted through a background-only fit of the data in each category. Background contributions from prompt leptons, including their shapes, are taken from MC simulations. In the case of non-prompt leptons the background shapes are taken from the Matrix Method. In the fit, the normalisation of  all backgrounds are allowed to vary within their uncertainties. 
The post-fit background yields are summarized in Table \ref{tab:Results:qqMWZPostFitYield} for the $q\bar{q}$ and VBF categories. The fit constrains the SM $WZ$ background estimate to the observed data, which reduces the total background uncertainty, pulling the modeling uncertainties by less than one standard deviation from their pre-fit values. None of the nuisance parameters are significantly pulled or constrained relative to their pre-fit values in the background-only fit.

\begin{figure}
\begin{center}
\subfloat[ ]{\includegraphics[width=0.5\textwidth]{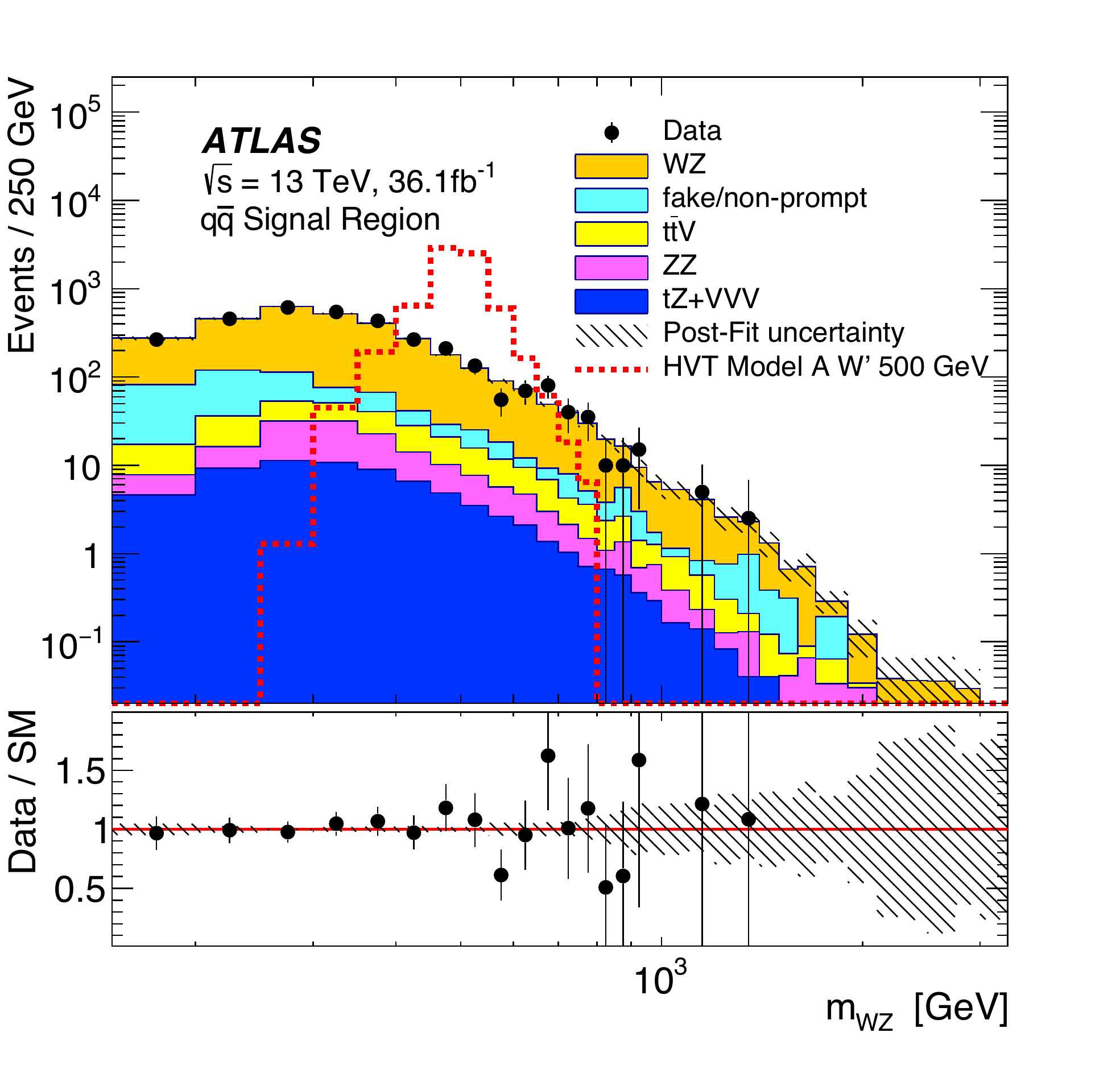}\label{SR_WZMassa}}
\subfloat[ ]{\includegraphics[width=.5\textwidth]{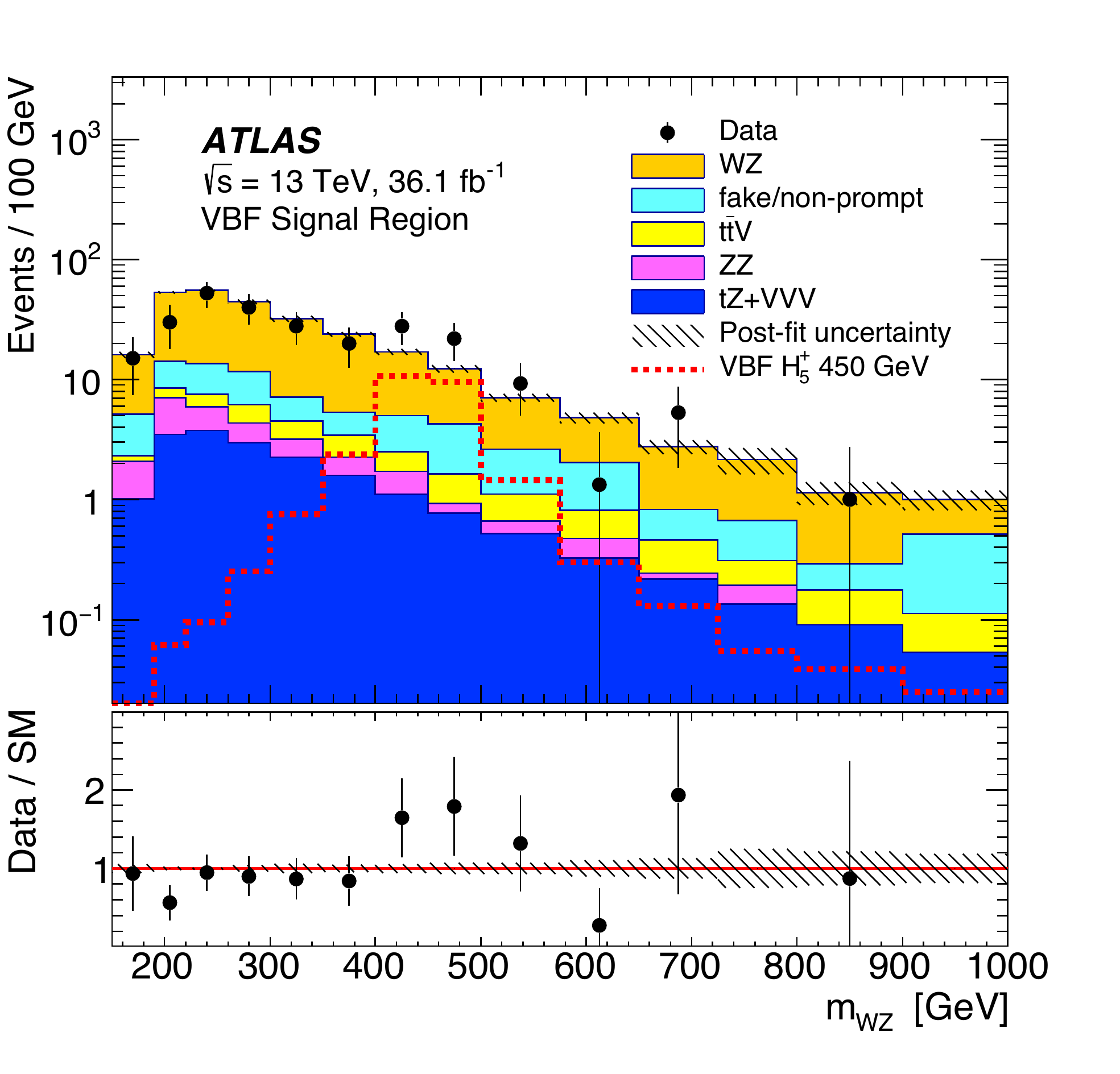}\label{SR_WZMassb}}
\caption{Observed and expected distributions of the $WZ$ invariant mass \protect\subref{SR_WZMassa} in the $q\bar{q}$ and \protect\subref{SR_WZMassb} in the VBF categories after applying all selection criteria. 
Signal predictions are overlaid, normalized to the predicted cross sections.  The uncertainty in the total background prediction, shown as shaded bands, combines statistical, theory and systematic contributions. The lower panel show the ratios of the observed data to the background predictions. \\
}
\label{SR_WZMass}
\end{center}
\end{figure}

Figure~\ref{SR_WZMass} shows the post-fit $m_{WZ}$ distribution for the $q\bar{q}$ and VBF categories. 
The largest difference between the observed data and the SM background prediction is in the VBF category. A local excess of events at a resonance mass of around $450~\GeV$ can be seen in Figure~\ref{SR_WZMassb}. The local significances for signals of $H_{5}^{\pm}$ and of a heavy vector $W'$ are 2.9 and 3.1 standard deviations, respectively. The respective global significances calculated using the Look Elsewhere method as in Ref.~\cite{ATL-PHYS-PUB-2011-011} and evaluated up to a mass of 900~\GeV, are 1.6 and 1.9 standard deviations. In the $q\bar{q}$ category the largest difference between the observed data and the SM background prediction is located around a mass of $700~\GeV$ with a local significance of 1.2 standard deviations.

Upper limits are set on the product of the production cross section of new resonances and their decay branching ratio into $WZ$.
Exclusion intervals are derived using the CL$_\text{s}$ method~\cite{Read:2002hq} in the asymptotic approximation~\cite{Cowan:2010js}.  For masses higher than 900 (700)~\GeV\ in $q\bar{q}$ (VBF)  category, the small number of expected events makes the asymptotic approximation imprecise and the limits are calculated using pseudo-experiments.
The limit set on the signal strength  $\mu$ is then translated into a limit on the signal cross section times branching ratio, $\sigma \times \mathcal{B}(W' \rightarrow WZ )$, using the theoretical cross section and branching ratio for the given signal model.

\begin{table}[th]
\begin{center}
\begin{tabular}{l|c|c }
\hline
\hline
           & $q\bar{q}$ Signal Region & VBF Signal Region\\ 
       \hline
      $WZ$                             & $521 \pm 29$              & $87 \pm 12$   \\
     Fake/non-prompt                   & $64 \pm 13$               & $15 \pm 4$  \\
     $t\bar{t}V$                       & $29 \pm 4$                & $4.9     \pm 0.8$    \\
     $ZZ$                              & $18.9 \pm 2.0$            & $4.4  \pm 1.0$   \\
     $tZ +VVV$                         & $14.1 \pm 2.9$            & $8.1 \pm 1.8$    \\
    \hline
  Total Background                     & $647 \pm 25$              & $120 \pm 11$ \\
     Observed                          & $650$                     & 114\\
     \hline
     \hline
\end{tabular}
\end{center}
\caption{Expected and observed yields in the $q\bar{q}$ and VBF signal regions. Yields and uncertainties are evaluated after a background-only fit to the data in the $q\bar{q}$ or VBF signal regions after applying all selection criteria.  The uncertainty in the total background estimate is smaller than the sum in quadrature of the individual background contributions due to anti-correlations between the estimates of different background sources.}
\label{tab:Results:qqMWZPostFitYield}
\end{table}

\begin{figure}[!htbp]
\begin{center}
\includegraphics[width=.7\textwidth]{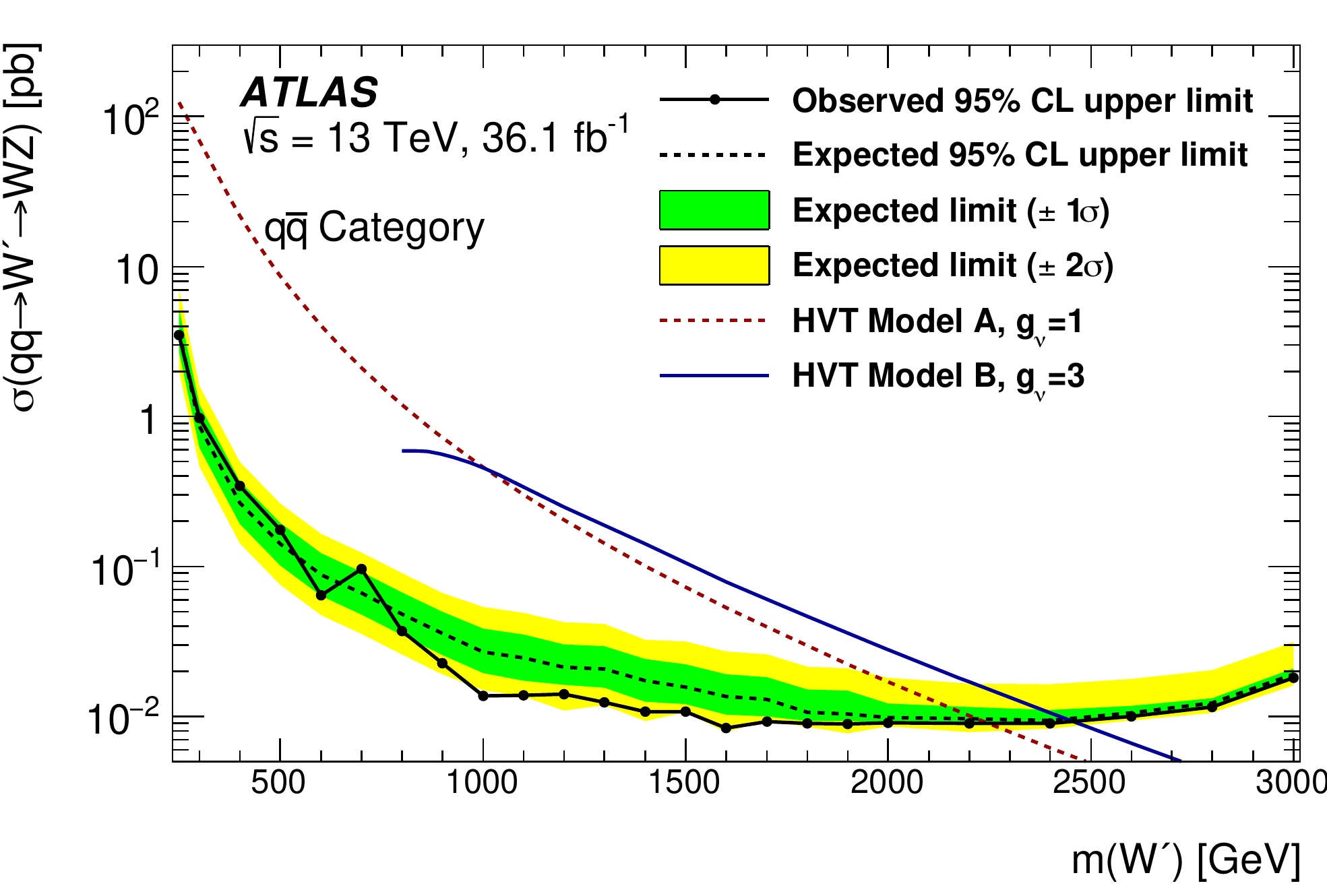}
\end{center}
 \caption{Observed and expected 95\% CL upper limits on $\sigma \times \mathcal{B}(W' \rightarrow W^{\pm}Z )$  for the $q\bar{q}$ production of a $W'$ boson in the HVT models as a function of its mass.  
 The theoretical predictions for HVT Models A with $g_V=1$ and B with $g_V = 3$ are also shown. 
 }
 \label{fig:Results:qqLimits}
 \end{figure}

\begin{figure}[!htbp]
\begin{center}
\includegraphics[width=.7\textwidth]{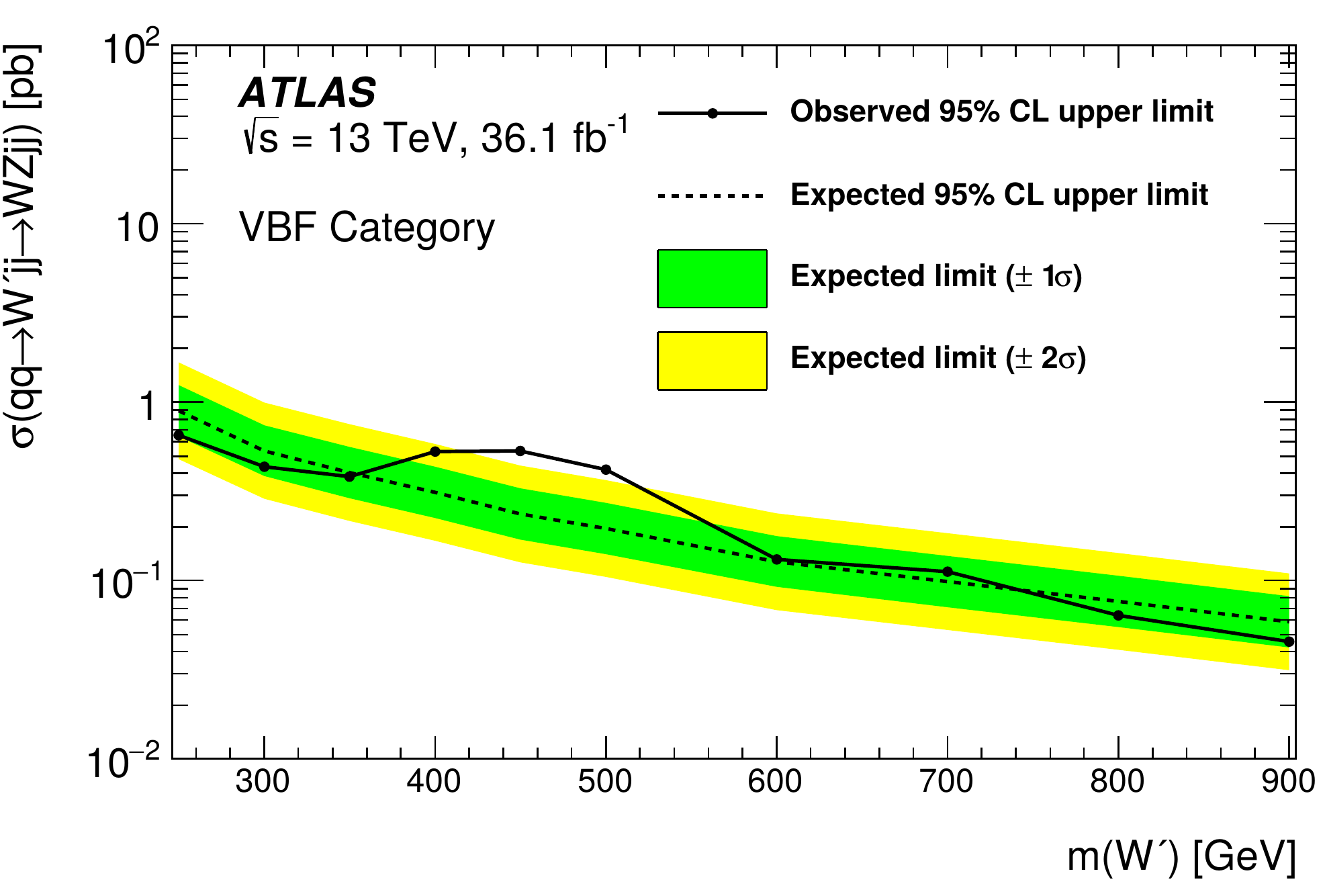}
\end{center}
 \caption{Observed  and expected  95\% CL upper limits on $\sigma \times \mathcal{B}(W' \rightarrow W^{\pm}Z )$  for the VBF production of a $W'$ boson in the HVT Model, with parameter $c_F = 0$, as a function of its mass.  
 }
\label{fig:Results:vbsHVTLimits}
\end{figure}

Figure \ref{fig:Results:qqLimits} presents the observed and expected limits on $\sigma \times \mathcal{B}(W' \rightarrow WZ )$ at 95\% CL for the HVT model in the $q\bar{q}$  category. Masses below $2260~\GeV$ can be excluded for Model A and $2460~\GeV$ for Model B. For resonance masses above $2~\TeV$ the exclusion limits become worse due to the acceptance losses at high mass.
For the VBF process, the limit on $\sigma \times \mathcal{B}(W' \rightarrow WZ )$ is shown in Figure~\ref{fig:Results:vbsHVTLimits}.

%----- Limits

\begin{figure}[!htbp]
\begin{center}
\subfloat[ ]{\includegraphics[width=.5\textwidth]{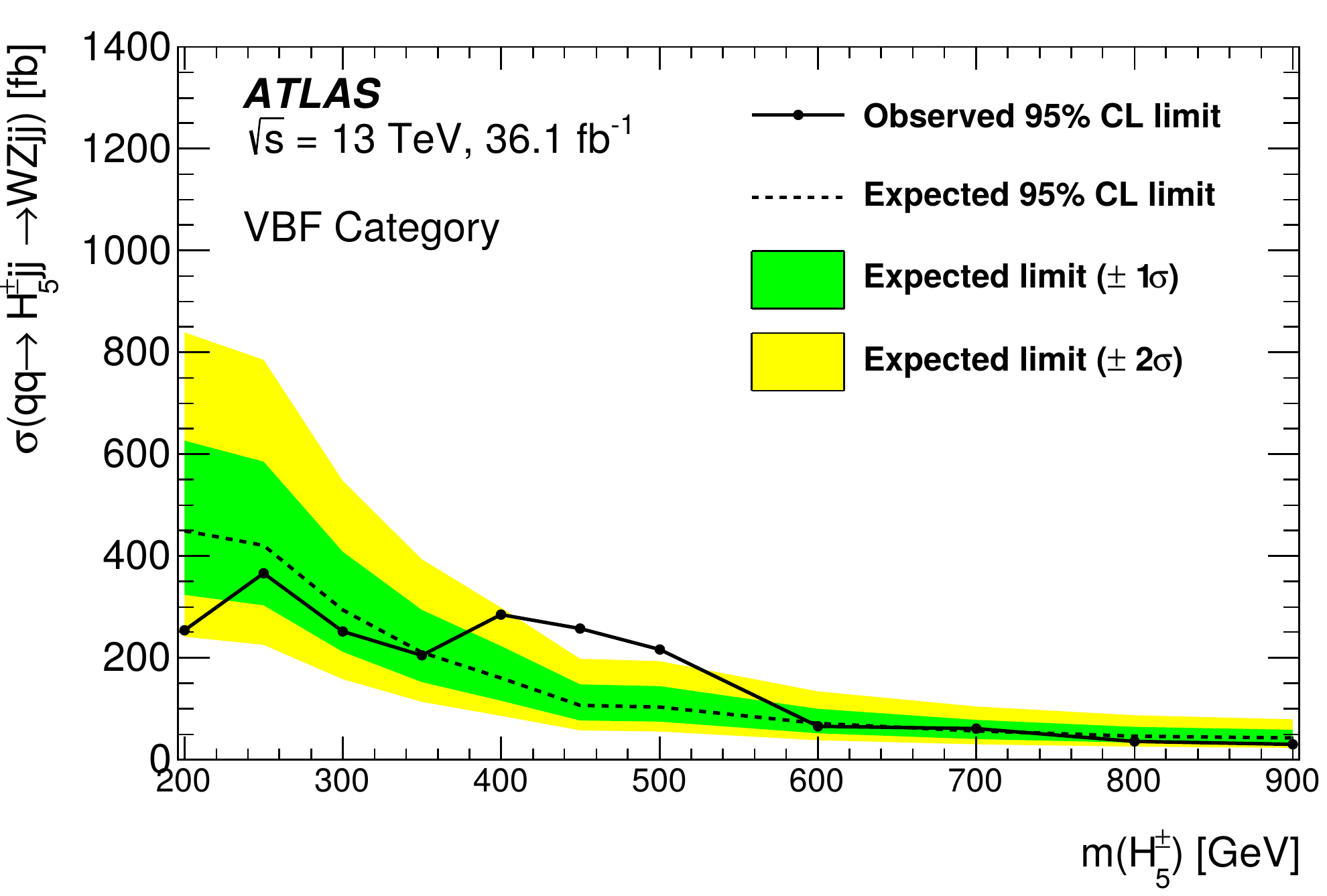}\label{fig:Results:H5Limitsa}}
\subfloat[ ]{\includegraphics[width=.5\textwidth]{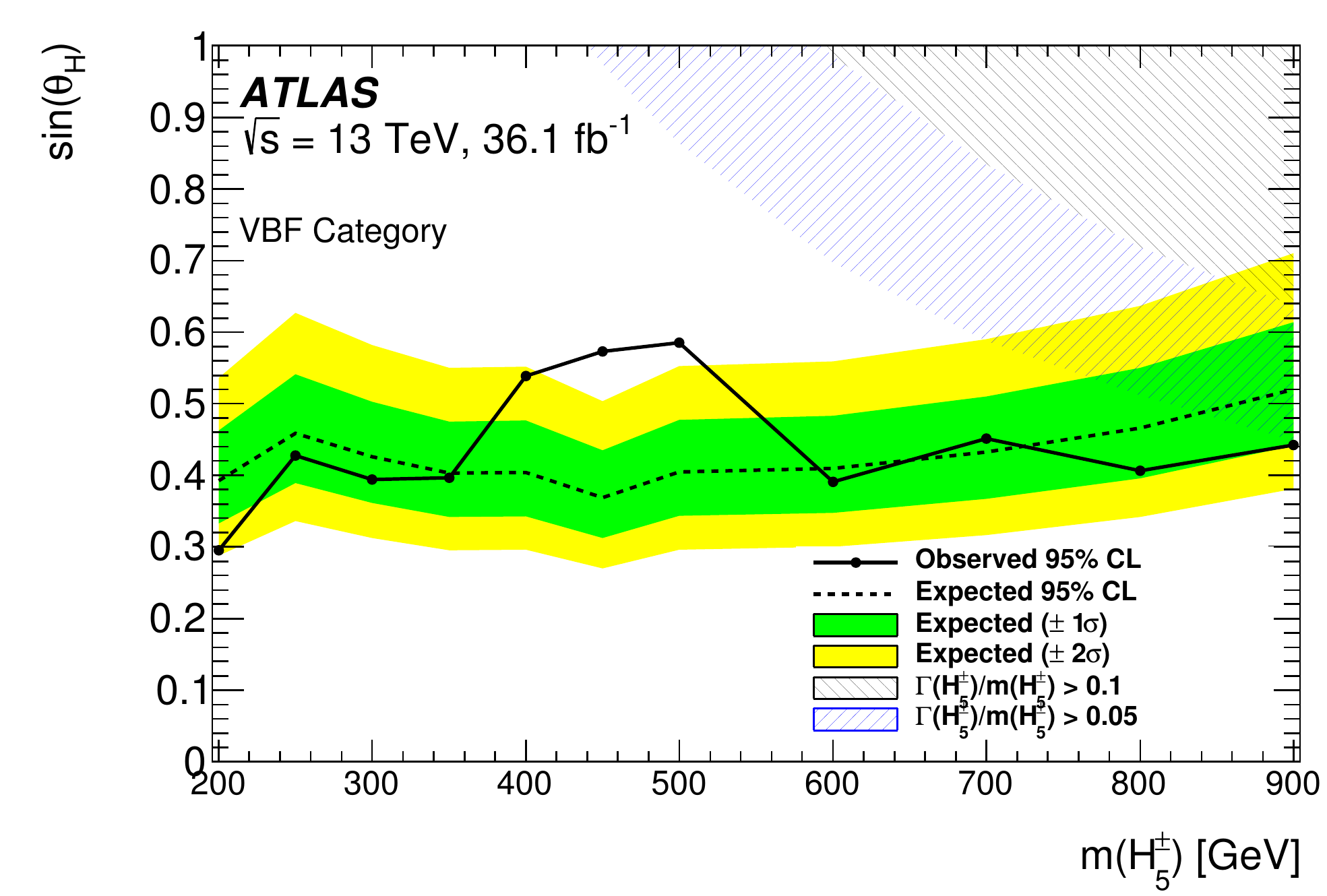}\label{fig:Results:H5Limitsb}}
\end{center}
 \caption{Observed  and expected  95\% CL upper limits on \protect\subref{fig:Results:H5Limitsa} $\sigma \times \mathcal{B}(H_{5}^{\pm} \rightarrow W^{\pm}Z )$  and  \protect\subref{fig:Results:H5Limitsb} the parameter $\sin\theta_H$ of the GM Model as a function of $m_{H_{5}^{\pm}}$. The shaded region shows where the theoretical intrinsic width of the resonance would be larger than 5\% or 10\% of the mass. 
 }
\label{fig:Results:H5Limits}
\end{figure}

Observed and expected exclusion limits at 95\% CL on $\sigma \times \mathcal{B}(H_{5}^{\pm} \rightarrow W^\pm Z )$ and on the mixing parameter $\sin\theta_H$ of the GM Model are shown in Figure \ref{fig:Results:H5Limits} as a function of $m_{H_{5}^{\pm}}$. The intrinsic width of the scalar resonance, for $\sin\theta_H =0.5$, is narrower than the detector resolution in the mass region explored. The shaded regions show the parameter space for which the $H_{5}^{\pm}$  width exceeds 5\% and 10\% of $m_{H_{5}^{\pm}}$.

%----- Limits HVT VBS                                                                                                  

%----- Limits H5                                                                                                       

%-------------------------------------------------------------------------------

% All figures and tables should appear before the summary and conclusion.
% The package placeins provides the macro \FloatBarrier to achieve this.
 \FloatBarrier

%-------------------------------------------------------------------------------
\section{Conclusion}
\label{sec:conclusion}
%-------------------------------------------------------------------------------
A search is performed for resonant $WZ$  production in fully leptonic final states  (electrons and muons) using 36.1~fb$^{-1}$ of $\sqrt{s} = 13~\TeV$ $pp$ data collected by the ATLAS experiment at the LHC during the 2015 and 2016 run periods. Two different production modes are considered using quark--antiquark annihilation and vector-boson fusion. 

The data in the $q\bar q$ fusion category are found to be consistent with Standard Model predictions.  The results are used to derive upper limits at 95\% CL on the cross section times branching ratio of the phenomenological Heavy Vector Triplet benchmark Model A (Model B) with coupling constant $g_V=1$ ($g_V=3$)
as a function of the resonance mass, with no evidence of heavy resonance production 
for masses below 2260 (2460)~\GeV. 

In the case of the VBF production processes, limits on the production cross section times branching ratio are obtained as a function of the mass of a charged member of a heavy vector triplet or of the fiveplet scalar in the Georgi--Machacek model. The results show a local excess of events over the Standard Model expectations at a resonance mass of around 450~\GeV. 
The local significances for signals of $H_{5}^{\pm}$ and of a heavy vector $W'$ boson are 2.9 and 3.1 standard deviations respectively. The respective global significances calculated considering the Look Elsewhere effect are 1.6 and 1.9 standard deviations respectively. 

%-------------------------------------------------------------------------------
\section*{Acknowledgements}
%-------------------------------------------------------------------------------
% Acknowledgements for papers with collision data
% Version 05-Oct-2018

% Standard acknowledgements start here
%----------------------------------------------
We thank CERN for the very successful operation of the LHC, as well as the
support staff from our institutions without whom ATLAS could not be
operated efficiently.

We acknowledge the support of ANPCyT, Argentina; YerPhI, Armenia; ARC, Australia; BMWFW and FWF, Austria; ANAS, Azerbaijan; SSTC, Belarus; CNPq and FAPESP, Brazil; NSERC, NRC and CFI, Canada; CERN; CONICYT, Chile; CAS, MOST and NSFC, China; COLCIENCIAS, Colombia; MSMT CR, MPO CR and VSC CR, Czech Republic; DNRF and DNSRC, Denmark; IN2P3-CNRS, CEA-DRF/IRFU, France; SRNSFG, Georgia; BMBF, HGF, and MPG, Germany; GSRT, Greece; RGC, Hong Kong SAR, China; ISF and Benoziyo Center, Israel; INFN, Italy; MEXT and JSPS, Japan; CNRST, Morocco; NWO, Netherlands; RCN, Norway; MNiSW and NCN, Poland; FCT, Portugal; MNE/IFA, Romania; MES of Russia and NRC KI, Russian Federation; JINR; MESTD, Serbia; MSSR, Slovakia; ARRS and MIZ\v{S}, Slovenia; DST/NRF, South Africa; MINECO, Spain; SRC and Wallenberg Foundation, Sweden; SERI, SNSF and Cantons of Bern and Geneva, Switzerland; MOST, Taiwan; TAEK, Turkey; STFC, United Kingdom; DOE and NSF, United States of America. In addition, individual groups and members have received support from BCKDF, the Canada Council, CANARIE, CRC, Compute Canada, FQRNT, and the Ontario Innovation Trust, Canada; EPLANET, ERC, ERDF, FP7, Horizon 2020 and Marie Sk{\l}odowska-Curie Actions, European Union; Investissements d'Avenir Labex and Idex, ANR, R{\'e}gion Auvergne and Fondation Partager le Savoir, France; DFG and AvH Foundation, Germany; Herakleitos, Thales and Aristeia programmes co-financed by EU-ESF and the Greek NSRF; BSF, GIF and Minerva, Israel; BRF, Norway; CERCA Programme Generalitat de Catalunya, Generalitat Valenciana, Spain; the Royal Society and Leverhulme Trust, United Kingdom.

The crucial computing support from all WLCG partners is acknowledged gratefully, in particular from CERN, the ATLAS Tier-1 facilities at TRIUMF (Canada), NDGF (Denmark, Norway, Sweden), CC-IN2P3 (France), KIT/GridKA (Germany), INFN-CNAF (Italy), NL-T1 (Netherlands), PIC (Spain), ASGC (Taiwan), RAL (UK) and BNL (USA), the Tier-2 facilities worldwide and large non-WLCG resource providers. Major contributors of computing resources are listed in Ref.~\cite{ATL-GEN-PUB-2016-002}.
%----------------------------------------------

%-------------------------------------------------------------------------------
%\clearpage
%\appendix
%\part*{Appendix}
%\addcontentsline{toc}{part}{Appendix}
%-------------------------------------------------------------------------------

%In a paper, an appendix is used for technical details that would otherwise disturb the flow of the paper.
%Such an appendix should be printed before the Bibliography.

%-------------------------------------------------------------------------------
% If you use biblatex and either biber or bibtex to process the bibliography
% just say \printbibliography here
\printbibliography
% If you want to use the traditional BibTeX you need to use the syntax below.
%\bibliographystyle{bib/bst/atlasBibStyleWoTitle}
%\bibliography{WZlvllPaper,bib/ATLAS,bib/CMS,bib/ConfNotes,bib/PubNotes}
%-------------------------------------------------------------------------------

%-------------------------------------------------------------------------------
% Auxiliary material - comment out the following line if you do not have any
%\include{WZlvllPaper-auxmat}
%\clearpage
%\appendix
%\part*{Auxiliary material}
%\addcontentsline{toc}{part}{Auxiliary material}
%\label{Aux}
%\input{Auxiliary}
%-------------------------------------------------------------------------------
\clearpage 
% ATLAS Collaboration author list
% Reference date of EXOT-2016-11 is 2018-01-26
% Author list last updated on date 25-JUL-18
% Data extracted on 25-Jul-2018 for paper reference EXOT-2016-11
% at 9:50am
 
\begin{flushleft}
{\Large The ATLAS Collaboration}

\bigskip

M.~Aaboud$^\textrm{\scriptsize 34d}$,    
G.~Aad$^\textrm{\scriptsize 99}$,    
B.~Abbott$^\textrm{\scriptsize 124}$,    
O.~Abdinov$^\textrm{\scriptsize 13,*}$,    
B.~Abeloos$^\textrm{\scriptsize 128}$,    
D.K.~Abhayasinghe$^\textrm{\scriptsize 91}$,    
S.H.~Abidi$^\textrm{\scriptsize 165}$,    
O.S.~AbouZeid$^\textrm{\scriptsize 143}$,    
N.L.~Abraham$^\textrm{\scriptsize 153}$,    
H.~Abramowicz$^\textrm{\scriptsize 159}$
% ...
\bigskip
\\

% ...
$^{13}$Institute of Physics, Azerbaijan Academy of Sciences, Baku; Azerbaijan.\\
% ...
$^{34}$$^{(a)}$Facult\'e des Sciences Ain Chock, R\'eseau Universitaire de Physique des Hautes Energies - Universit\'e Hassan II, Casablanca;$^{(b)}$Centre National de l'Energie des Sciences Techniques Nucleaires (CNESTEN), Rabat;$^{(c)}$Facult\'e des Sciences Semlalia, Universit\'e Cadi Ayyad, LPHEA-Marrakech;$^{(d)}$Facult\'e des Sciences, Universit\'e Mohamed Premier and LPTPM, Oujda;$^{(e)}$Facult\'e des sciences, Universit\'e Mohammed V, Rabat; Morocco.\\
% ...
$^{91}$Department of Physics, Royal Holloway University of London, Egham; United Kingdom.\\
% ...
$^{99}$CPPM, Aix-Marseille Universit\'e, CNRS/IN2P3, Marseille; France.\\
% ...
$^{124}$Homer L. Dodge Department of Physics and Astronomy, University of Oklahoma, Norman OK; United States of America.\\
% ...
$^{128}$LAL, Universit\'e Paris-Sud, CNRS/IN2P3, Universit\'e Paris-Saclay, Orsay; France.\\
% ...
$^{143}$Santa Cruz Institute for Particle Physics, University of California Santa Cruz, Santa Cruz CA; United States of America.\\
% ...
$^{153}$Department of Physics and Astronomy, University of Sussex, Brighton; United Kingdom.\\
% ...
$^{159}$Raymond and Beverly Sackler School of Physics and Astronomy, Tel Aviv University, Tel Aviv; Israel.\\
% ...
$^{165}$Department of Physics, University of Toronto, Toronto ON; Canada.\\
% ...
$^{a}$ Also at  Department of Physics, University of Malaya, Kuala Lumpur; Malaysia.\\
$^{b}$ Also at Borough of Manhattan Community College, City University of New York, NY; United States of America.\\
$^{c}$ Also at Centre for High Performance Computing, CSIR Campus, Rosebank, Cape Town; South Africa.\\
$^{d}$ Also at CERN, Geneva; Switzerland.\\
$^{e}$ Also at CPPM, Aix-Marseille Universit\'e, CNRS/IN2P3, Marseille; France.\\
% ...
$^{*}$ Deceased

\end{flushleft}

% Created with Glance <Atlas.Glance@cern.ch>

%-------------------------------------------------------------------------------
% Extra tables etc. for HepData - comment in the following line if you have any
% \include{WZlvllPaper-hepdata}
%-------------------------------------------------------------------------------

\end{document}